%% file: conference_itw2025.tex
\documentclass[conference]{IEEEtran}
\IEEEoverridecommandlockouts
\usepackage{cite}
\usepackage{amsmath,amssymb,amsfonts}
\usepackage{algorithmic}
\usepackage{graphicx}
\usepackage{textcomp}
\usepackage{xcolor}
\def\BibTeX{{\rm B\kern-.05em{\sc i\kern-.025em b}\kern-.08em
    T\kern-.1667em\lower.7ex\hbox{E}\kern-.125emX}}

\def\*#1{\mathbf{#1}}
\newcommand{\vect}[1]{{\mathbf{#1}}}
\newcommand{\mat}[1]{{\mathbf{#1}}}

\newcommand{\TS}[1]{\mathcal{T}_{#1}}

\renewcommand{\sf}[1]{\mathsf{#1}}

\newcommand{\balpha}{\boldsymbol{\alpha}}

\newcommand{\yue}[1]{{\color{brown}#1}}
\renewcommand{\yue}[1]{{#1}}

\newcommand{\cor}[1]{{\color{blue}#1}}

\renewcommand{\cor}[1]{{#1}}

\newcommand{\lo}[1]{{#1}}

\newcommand{\gc}[1]{{#1}} 
\renewcommand{\gc}[1]{{\color{green}#1}}
\renewcommand{\gc}[1]{}
\newif\ifADDpagenumber
\ADDpagenumbertrue

\allowdisplaybreaks[4]
\usepackage{amsthm}
\newtheorem{theorem}{Theorem}
\newtheorem{lemma}{Lemma}
\newtheorem{corollary}{Corollary}
\newtheorem{remark}{Remark}

\begin{document}

\title{Interference Alignment for Multi-cluster Over-the-Air Computation}


\author{
    \IEEEauthorblockN{Lucas Sempéré\IEEEauthorrefmark{1}, Yue Bi\IEEEauthorrefmark{2}, Yue Wu\IEEEauthorrefmark{3}, Pengwenlong Gu\IEEEauthorrefmark{4} ,Selma Boumerdassi\IEEEauthorrefmark{4}}
    

    \IEEEauthorblockA{\IEEEauthorrefmark{1} \textit{Shanghai Paris Elite Institute of Technology, Shanghai Jiao Tong University}, China;\;}
    \IEEEauthorblockA{\IEEEauthorrefmark{2} \textit{LTCI, Telecom Paris, IP Paris}, 91120 Palaiseau, France; \;}
    \IEEEauthorblockA{\IEEEauthorrefmark{3}\textit{School of Electronic Information and Electrical Engineering, Shanghai Jiao Tong University}, China;\;}

    \IEEEauthorblockA{\IEEEauthorrefmark{4}\textit{CEDRIC, CNAM, France}}

    \IEEEauthorblockA{\{lucas.sempere, bi\}@telecom-paris.fr, wuyue@sjtu.edu.cn, \{gup, selma.boumerdassi\}@cnam.fr}
}

\maketitle

\ifADDpagenumber
\thispagestyle{plain} 
\pagestyle{plain}
\fi

\begin{abstract}
One of the main challenges facing Internet of Things (IoT) networks is managing interference caused by the large number of devices communicating simultaneously, particularly in multi-cluster networks where multiple devices simultaneously transmit to their respective receiver. Over-the-Air Computation (AirComp) has emerged as a promising solution for efficient real-time data aggregation, yet its performance suffers in dense, interference-limited environments. To address this, we propose a novel Interference Alignment (IA) scheme tailored for up-link AirComp systems. Unlike previous approaches, the proposed method scales to an arbitrary number $\sf K$ of clusters and enables each cluster to exploit half of the available channels, instead of only $\tfrac{1}{\sf K}$ as in time-sharing. In addition, we develop schemes tailored to scenarios where users are shared between adjacent clusters.
\end{abstract}


\section{Introduction}\label{sec:Intro}

The Internet of Things (IoT) is rapidly expanding, with billions of devices like sensors and smartphones generating massive data at the network edge \cite{zhu_one_bit_aggregation_2021}. Applications in smart cities, healthcare, wild-area monitoring, and autonomous driving rely on these data for real-time decisions. As devices multiply, managing data efficiently becomes a key challenge. Efficient processing mechanisms are essential for reliable IoT performance.

Over-the-Air Computation (AirComp) has emerged as a crucial technique for efficient wireless data aggregation in massive IoT networks. AirComp exploits the superposition property of wireless channels to directly compute functions of distributed data, thereby enabling over-the-air aggregation. \lo{Unlike traditional orthogonal multi-access schemes, all devices transmit their data simultaneously over the full set of radio resources rather than a fraction of them. As a result, the signal of each transmitter within the same cluster is added over the air and arrive at the receivere as an aggregated sum weighted by the channel coefficients \cite{zhu_one_bit_aggregation_2021}. }
This approach significantly reduces communication overhead and latency, making it particularly suitable for industrial IoT \cite{zhu_mimo_aircomp_2019}. However, a key challenge for AirComp lies in managing interference in wireless networks. Traditional methods fail to scale efficiently as the number of devices increases, leading to errors in data aggregation. Cooperative interference management frameworks has been studied to minimize the mean squared error (MSE) in aggregated signals \cite{zhu_aircomp_aggregation_2021}. Along with optimized power control, beamforming has been the main strategy when it comes to managing interference \cite{wen_aircomp_2019, li_aircomp_2019}. 
Meanwhile, another technique, Interference Alignment (IA), offers the potential to further improve AirComp performance in multi-user environments, though its integration into such systems remains largely unexplored.

\lo{
IA has marked a significant breakthrough in wireless communications by enabling the efficient management of interference. This approach optimizes the use of transmission resources and improves network capacity by concentrating information flows into interference-free signal subspaces. By skillfully pre-encoding the signals, interference can be aligned into a smaller subspace, allowing a larger portion of the channel to be used for transmitting desired signals. In other words, interference signals are intentionally overlapped through encoding, and their subspace is separated from that of the useful signal to preserve decodability. IA schemes have been shown to be operating close to the theoretical capacity limit of a wireless channel \cite{jafar_interference_2011}. Initially introduced for the $\sf K$-user interference channel, IA enables each user to access half of the signal space simultaneously \cite{cadambe_interference_2008}. This technique has been applied to various channel configurations, such as the X-channel \cite{cadambe_interference_2009}, cooperative or coordinated channels \cite{annapureddy_DoF_2012, bi_DoF_2022}, and MIMO multi cluster networks \cite{tang_interference_2013, mohammadghasemi_feedback_2019}. In all of these cases, IA has proven to be efficient, helping achieve higher degrees of freedom and thereby improving network capacity.}
\gc{IA is a major breakthrough in wireless communications, enabling efficient interference management by aligning unwanted signals into a reduced subspace and preserving more channel dimensions for desired signals. This approach, initially developed for the $\sf K$-user interference channel where each user accesses half of the signal space \cite{cadambe_interference_2008}, has since been applied to X-channels \cite{cadambe_interference_2009}, cooperative/coordination setups \cite{annapureddy_DoF_2012, bi_DoF_2022}, and MIMO multicell networks \cite{tang_interference_2013, mohammadghasemi_feedback_2019}. IA schemes operate close to the channel’s capacity limit \cite{jafar_interference_2011} and significantly improve network degrees of freedom and capacity.}

Motivated by the strengths of both approaches, recent studies have begun exploring the integration of IA with AirComp. For example, the SIA scheme proposed in \cite{lan_combined_2020} enables IA in AirComp systems, but only for two-cluster scenarios. The CRDIA scheme in \cite{li_interference_2024} addresses interference in multi-cluster networks but relies solely on numerical evaluations without providing analytical insights.
\yue{In contrast, this work provides the first analytical results for an IA–AirComp framework in multi-cluster networks with an arbitrary number of possibly overlapping clusters, supported by a theoretical analysis that establishes its scalability and performance.}

Formally, we consider a network with $\sf K$ clusters, each comprising a group of $\sf r$ transmitters (Txs) and a single receiver (Rx). Unlike most existing works, where clusters are disjoint, we allows clusters to share Txs, such that the signals of some Txs are intended for multiple Rxs.  Txs in each cluster aim to deliver the sum of their local data to their associated Rx. Inspired by \cite{cadambe_interference_2008, bi_wireless_2024}, we jointly design precoding matrices by integrating IA with AirComp. \yue{We show that useful signals can be separated from interference under our scheme. Unlike conventional constructions of precoding matrices, which require the channel matrices of the desired signals to be independent from both the interference channel matrices and the precoding matrices, our construction leverages the parity of the exponents to guarantee separation.}
Our proposed scheme achieves a performance gain of a factor of $\sf K/2$ compared to the conventional TDMA-AirComp approach when the number of Txs shared between adjacent clusters is less than one, and $\sf K/3$ when more Txs are shared.

\textit{Notations:}{ We use sans-serif font for constants, boldface for vectors and matrices, and calligraphic font for sets. \lo{The sets of complex numbers and natural numbers are denoted by $\mathbb{C}$ and $\mathbb{N}$.} For a finite set $\mathcal{A}$, $|\mathcal{A}|$ denotes its cardinality. For any $n \in \mathbb{N}$, we define $[\![1,n]\!] \triangleq \{1,2,\dots,n\}$. Let $\textbf{1}$ denote the all-ones vector, with dimensions determined by context. For any matrix $\mat{B}$, we write $\mat{B}^{-1}$ for its inverse (when full rank), $\mat{B}^{T}$ for its transpose, $\det(\mat{B})$ for its determinant, and $\text{span}(\mat{B})$ for the column space it spans. For a vector $\vect{X}$ of size $n$, $X(i)$ with $i \in [\![1,n]\!]$ denotes its $i$-th component. Similarly, for an $n \times m$ matrix $\mat{B}$, $B(i,j)$ with $(i,j) \in [\![1,n]\!] \times [\![1,m]\!]$ denotes its $(i,j)$-th entry. Finally, the notation $[\vect{s}{i} : i \in \mathcal{S}]$ refers to the matrix whose columns are the vectors in ${\vect{s}{i}}_{i \in \mathcal{S}}$.}

\input{contents/setup}

\input{contents/main_res}
\input{contents/proof}

\section{Conclusion}\label{sec:ccl}

This work introduces an advancement in multi-cluster network design for AirComp systems through a specially designed IA scheme. Specifically, it establishes a lower bound on the A-SDoF of $\frac{\sf K}{2}$ for an arbitrary number of clusters, $\sf K$, while remaining effective even when adjacent clusters share a single Tx. Furthermore, for the case of multiple shared Txs, we propose an alternative scheme achieving an A-SDoF of $\frac{\sf K}{3}$. Our results demonstrate the potential for substantial performance improvements, allowing all $\sf K$ Rxs to receive simultaneously, unlike traditional AirComp and IA schemes, which are limited to one receiver or require resource sharing per group. 

\lo{In future work, we aim to present results for scenarios with a limited data size. While analytical treatment in these cases is challenging, we will provide simulations and numerical evaluations to support our findings. For tractability, this paper focuses on a one-dimensional network; extensions to settings with two-dimensional topologies (e.g., hexagonal networks) are left for future work, as they can be addressed using similar methods.}


\section*{Acknowledgement}
We thank P. Martins and X. Xin for helpful discussions. This work has been supported by National Key R\&D Program of China under Grant No 2023YFB2704903.

\bibliographystyle{IEEEtran}
\bibliography{IEEEabrv,references}

\lo{
\appendix
\input{contents/Lemma_full_rank}

}

\end{document}

%% file: contents/setup.tex
\section{Channel model}\label{sec:channel_model}

Consider a network with $\sf K$ \yue{cluster. Cluster} $\ell$ contains a Tx group $\TS{\ell} \subset [\![1, \sf K]\!]$ with $\sf r$ Txs and a unique Rx. 
\yue{We assume each cluster has 2 neighbors, except cluster 1 and $\sf K$, at the extremities with only one neighbor. We allow two adjacent clusters to share Txs, contributing to both messages simultaneously. }
\yue{
We define
\begin{equation}
    \sf r_{\ell-1,\ell} = |\TS{\ell-1} \cap \TS{\ell}| \in [\![1, \sf r]\!]  \quad \forall \ell \in [\![1, \sf K]\!]
\end{equation}
as the number of Txs shared between the Tx groups $\ell-1$ and $\ell$.
For instance, $\sf r_{\ell-1,\ell}=1$ indicates that groups $\ell-1$ and $\ell$ share exactly one transmitter, while $\sf r_{\ell-1,\ell}=0$ means the two groups are independent. By convention, we set $\sf r_{0,1}=0$ since group $0$ does not exist. Moreover, we assume one Tx can be shared by at most 2 groups, which leads to the natural constraint we have $\sf r_{\ell-1,\ell} + \sf r_{\ell,\ell+1}\leq \sf r$ $\forall \ell \in [\![1, \sf K]\!]$.
We introduce $\sf r_{\ell} \triangleq \sum_{k=1}^{\ell} \sf r_{k-1,k}$ for all $\ell$ in $[\![1, \sf K]\!]$ that count from cluster $1$ to cluster $\ell$ the amount of Txs that are part of two clusters. 
Therefore, the total number of Txs $\sf M = \sf K \sf r - \sf r_{\sf K}$, and we label the group of Txs within a cluster by the given set:
\begin{equation*}
    \TS{\ell} = \{(\ell-1)\sf r +1 - \sf r_{\ell}, \ldots, \ell \sf r - \sf r_{\ell} \}, \quad \ell \in[ \![1, \sf K]\!]
\end{equation*}
The elements of this set will be called $t_{\ell,i}$ with $i \in [ \![1,\sf r]\!]$. }

\begin{figure}[ht]
    \centering
    \includegraphics[width=0.5\textwidth, height=0.4\textheight, keepaspectratio, clip]{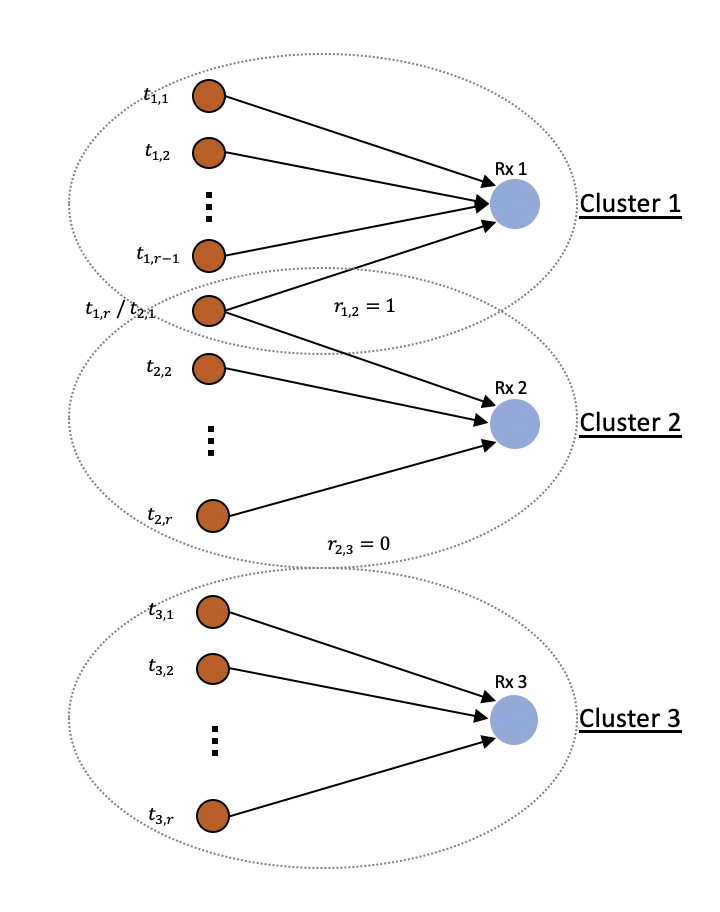}
    \caption{Illustration of the clusters $1$, $2$ and $3$ where $r_{1,2}=1$ and $r_{2,3}=0$. Each clusters is composed of $\sf r$ Txs and cluster $1$ and $2$ are sharing one of them.}
    \label{fig:example_setup}
\end{figure}

\yue{Following the classic AirComp channel model in~\cite{Sahin_survey_AirComp_2023} and in ~\cite{Nazer_compute_2011}, we consider that Tx $q \in [\sf M]$ transmit a length $\sf L$ vector $\vect{w}_q \in \mathbb{F}_p^{\sf L}$ whose entries is drawn independently and uniformly from $\mathbb{F}_p$. Rx~$\ell$ aims to recover the following modulo-$p$ sums of received messages:
\begin{equation}
    h\left(\left\{\vect{w}_{q} \colon q \in \TS{\ell}\right\}\right) = \bigoplus_{q \in \TS{\ell}} \vect{w}_{q}.
\end{equation}
Notice that this framework can be generalized to more complex computation tasks as in \cite{Jeon_computation_2014} and to cases involving nomographic functions, although these extensions are beyond the scope of this paper.}



Therefore, the Txs within a cluster are all synchronized and send their message simultaneously to their paired receiver. We assume that all clusters interfere with each other.

Each Rx $\ell$ observes a linear combination of signals of blocklength $\sf T$ sent by all Txs, corrupted by Gaussian noise. Denoting Tx~$q$'s input by $\vect{X}_{q} \in \mathbb{C}^{\sf T \times 1}$, Rx~$\ell$'s output by $\vect{Y}_{\ell} \in \mathbb{C}^{\sf T \times 1}$, $\mat{H}_{\ell,q} \in \mathbb{C}^{\sf T \times \sf T}$ the channel coefficient matrices and $\vect{Z}_{\ell} \in \mathbb{C}^{\sf T \times 1}$ the noise vector, the input-output relation of the network is:
\begin{IEEEeqnarray}{rCl}\label{eq:gen}
   \vect{Y}_\ell = \sum_{q=1}^{\sf M} \mat{H}_{\ell,q} \vect{X}_{q} + \vect{Z}_{\ell} , \quad \ell \in[ \![1, \sf K]\!],
\end{IEEEeqnarray}
where the complex-valued channel coefficient matrices $\mat{H}_{\ell,q}$ are diagonal, with diagonal entries drawn independently according to a bounded continuous distribution $[-\sf H_{\max},\sf H_{\max}]$. The standard circularly symmetric Gaussian noise vectors $\vect{Z}_{\ell}$ are also taken independently and identically distributed. Also, we assume knowledge to be casual and globally available, all $\mat{H}_{\ell,q}$ matrices are known by all terminals even before communication starts. 


\yue{The message transmitted by Tx $q$ is encoded into a vector $\vect{X}_{q} \in \mathbb{C}^{\sf T}$ with encoding function $f_q$. Therefore, the input signal of Tx $q$ is
\begin{equation}
    \vect{X}_{q} = [X_{q}(1), \cdots, X_{q}(\sf T)]^{\text{$T$}}=f_{q}(\vect{w}_{q})
\end{equation}}
The inputs need to satisfy the block-power constraint

\begin{equation}
    \frac{1}{\sf T} \sum_{t=1}^{\sf T}\mathbb{E}[|X_{q}(t)|^{2}] \leq \sf P, \quad q \in [\![1,\sf M]\!]
\end{equation}
given a power $\sf P > 0$.

Each Tx of the cluster $\ell$ sends its data block simultaneously to the Rx in the cluster. The Rx decodes the modulo-$p$ sum from the superposed signal $\vect{\mathbb{C}}^{\sf T}$ vector $\vect{Y}$ with a reliable decoder, so that: 
\begin{equation}\label{eq:estimates}
    \hat{h}\left(\left\{\vect{w}_{q} \colon q \in \TS{\ell}\right\}\right) = g_{\ell}(\vect{Y}_{\ell})
\end{equation}

We define the computational rate $\sf R_{\ell} = \frac{\sf L}{\sf T}\log_{2}(p)$ at Rx $\ell$ as the one in \cite{Nomo_functions_lattices}.
The capacity region $\mathcal{C}(\sf P)$ is defined as the set of all rate tuples $( \sf R_{\ell} : \ell \in [\![1,\sf K]\!] )$ achievable,
which means for any rate tuple in $\mathcal{C}(\sf P)$, and for any blocklength $\sf T$ there exist encoding functions $\{f_{q}\}_{q \in [\![1,\sf M]\!]}$ as described above and appropriate linear decoding functions $\{g_{\ell}\}_{\ell \in [\![1,\sf K]\!]}$ producing the estimates in \eqref{eq:estimates} so that the sequence of error probabilities
\begin{equation}
    p(error) \triangleq \textbf{Pr} \left[\bigcup_{\ell=1}^{\sf K} (\hat{h}_{\ell}(\{d_{q}\}_{q \in \TS{\ell}}) \neq h_{\ell}(\{d_{q}\}_{q \in \TS{\ell}}))\right]
\end{equation}
tends to $0$ as the blocklength $\sf T \to \infty$.

We define the AirComp-Sum Degrees of Freedom (A-SDoF) as the ratio between the achievable sum rate of the channel and the reference rate $\sf R$ of a single-cluster, one-dimensional channel, in the high SNR regime. Formally, it is given by:
\begin{equation}\label{eqn:a_sdof}
    \text{A-SDoF} \triangleq \lim_{\sf P \to \infty} \sup_{\vect{R} \in \mathcal{C}(\sf P)} \sum_{\ell = 1}^{\sf K} \frac{\sf R_{\ell}}{\sf R},
\end{equation}

\begin{remark}
In classical multi-user channels, such as interference channels or X-channels, the SDoF is defined as the ratio between the achievable sum rate and $\log \sf P$ in the high SNR regime, where $\log \sf P$ corresponds to the rate of a one-dimensional point-to-point channel. However, this traditional DoF definition cannot be directly applied to our AirComp-based scheme, where the goal is to compute a function of the transmitted messages rather than recover each message individually. 
Therefore, we define A-SDoF, as given in~\eqref{eqn:a_sdof}, by following the same principle. This metric serves as an analog of the classical SDoF, but adapted to AirComp systems.
A similar metric has been used in \cite{lan_combined_2020} to demonstrate the performance gain of IA in a two-cluster network. 
\end{remark}

%% file: contents/main_res.tex
\section{Main Results}\label{sec:main}

The main result of this paper is a new lower bound on the A-SDoF of the multi-cluster network described in the previous Section \ref{sec:channel_model}.

\yue{
\begin{theorem}\label{theorem1}
    The A-SDoF of the network described in Section \ref{sec:channel_model} is lower bounded as:
    \begin{equation}
        \textnormal{A-SDoF} \geq \frac{\sf K}{2} \quad \text{if} \quad \max_{\ell} \{ \sf r_{\ell-1, \ell} \} \leq 1
    \end{equation}
    \begin{equation}
        \textnormal{A-SDoF} \geq \frac{\sf K}{3} \quad \text{if} \quad \max_{\ell} \{\sf r_{\ell-1, \ell} \} \geq 2
    \end{equation}
\end{theorem}
}

\begin{IEEEproof}
We demonstrate the achievability of the lower bound by introducing a new IA scheme specifically designed for the AirComp system. A simple illustrative example is provided in Section~\ref{sec:proof_ex}, and the detailed proof is presented in Section~\ref{sec:proof_general}.
\end{IEEEproof}

\begin{corollary}\label{cor:AirComp_only}
In contrast with the A-SDoF found in Theoreme~\ref{theorem1}, the lower bound of the A-SDoF for the same network as described in Section \ref{sec:channel_model} but using a encoding scheme with only AirComp is $1$.
\end{corollary}

\begin{IEEEproof}
In a multi-cluster AirComp scheme, interference management is essential—without it, extracting the desired function becomes infeasible. A straightforward solution is to use a time-sharing strategy across the clusters , thereby eliminating inter-cluster interference. However, this requires dividing the transmission time into $\sf K$ separate slots, one for each cluster. As a result, each cluster operates only a fraction $\frac{1}{\sf K}$ of the time. According to the definition in \eqref{eqn:a_sdof}, we obtain a per-cluster DoF of $\frac{1}{\sf K}$ and a total Sum-DoF of 1.
\end{IEEEproof}

\lo{
\begin{remark}\label{cor:IA_only}
At the same time, for the same network, the lower bound of the Sum-DoF with an encoding scheme using only IA is $\frac{\sf K}{2 \sf r}$.
    
For IA scheme only, since each Rx needs to recover the sum of the data send by their Txs, we can again introduce a TDMA scheme but across the Txs within each cluster. To be clear, we divide the time into $\sf r$ time slot so that, for any time unit, only one Tx of each cluster is active. Therefore, at each time slot, the channel considered is an $\sf K$ interference channel. This channel have already been well studied by \cite{cadambe_interference_2008}. The Sum-DoF is $\frac{\sf K}{2}$ as mentioned in Section \ref{sec:Intro}. So the Sum-DoF of the entire network is $\frac{\sf K}{2 \sf r}$.
\end{remark}
}

The result in Corollary \ref{cor:AirComp_only} serves as a baseline, highlighting the inefficiency of naïve time-division approaches. A higher A-SDoF can be achieved by applying interference alignment while preserving the functional computation goals of AirComp systems.

\begin{remark}
    We recover the same results as in~\cite{lan_combined_2020} by considering the case of two clusters ($\sf K = 2$), each containing an arbitrary number $\sf r$ of Txs. The result extends to an arbitrary number of clusters and holds in the case of one shared Tx between two adjacent clusters.
\end{remark}

%% file: contents/proof.tex
\section{Degree of Freedom For the Multi-Cluster Overlapping Network}\label{sec:proof}

The main insight of this paper is how the idea of Interference alignment can be combined with an Over-the-Air setup in a network designed with overlapping clusters. We present an example with a small number of transmitters and receivers to introduce the key encoding. 

\subsection{Example scheme: $\sf K=3$, $\sf r=2$ and $\sf r_{1,2}= \sf r_{2,3}=1$}\label{sec:proof_ex}
Consider $\sf K=3$, $\sf r=2$ multi-cluster network \lo{as shown in Fig.~\ref{fig:example_scheme}}. To clarify, we consider a network with 3 clusters , each containing 2 Txs and 1 Rx. With our previous notation, we set $\sf r_{1,2} = \sf r_{2,3}=1$, meaning that clusters 1 and 2 share a common transmitter, as do clusters 2 and 3. Therefore, cluster 1 contains Tx 1 and 2, cluster 2 contains Tx 2 and 3, and cluster 3 contains Tx 3 and 4. Formally:
$$\TS{1} = \{1,2\}, \quad \TS{2} = \{2,3\}, \quad \TS{3}= \{3,4\}. $$
Therefore, Txs 2 and 3 are part of two different groups and send messages to both Rxs simultaneously. 

\lo{
\begin{figure}[ht]
    \centering
    \includegraphics[width=0.45\textwidth, trim=305 600 305 600, clip]{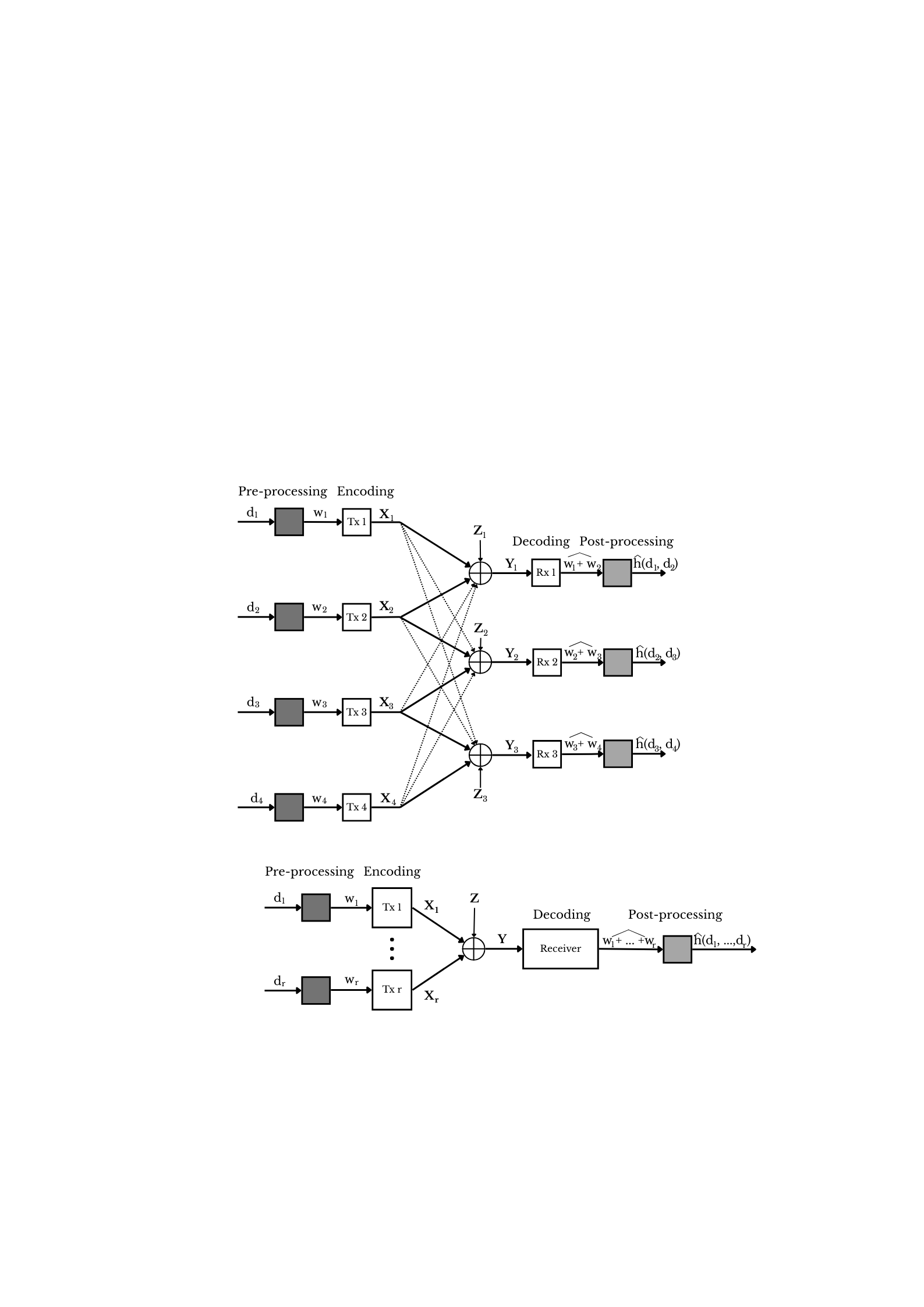} 
    \caption{Integration of a multi-cluster network into an AirComp transmission scheme for the Section \ref{sec:proof_ex}: $\sf K = 3$, $\sf r=2$. The plain arrows in the transmission part are the desired signals for the Rxs, and the dashed ones represent interference.}
    \label{fig:example_scheme}
\end{figure}
}

At the $q$-th Tx, the input signal is
\begin{equation}
    \vect{X}_{q} = \mat{C}_{q} \mat{V} \vect{x}_{q}
\end{equation}
\cor{with $\vect{x}_{q}$ being the modulated message $w_{q}$ and $\mat{C}_{q}$ and $\mat{V}$ being the pre-coding matrices to address AirComp and IA constraints respectively.} We will provide a detailed explanation of how to choose them later in this subsection.

The received signal at the $\ell$-th Rx can be written as:
\begin{multline}\label{eq:ex_Rxl}
    \mat{Y}_{\ell} = \mat{H}_{\ell,1} \mat{C}_{1} \mat{V} \vect{x}_{1} + 
                \mat{H}_{\ell,2} \mat{C}_{2} \mat{V} \vect{x}_{2} \\
                + \mat{H}_{\ell,3} \mat{C}_{3} \mat{V} \vect{x}_{3} + 
                \mat{H}_{\ell,4} \mat{C}_{4} \mat{V} \vect{x}_{4} + \mat{Z}_{\ell}
\end{multline}

We begin by examining the received signal at Rx 1, the desired codeword is the sum $\vect{x}_{1} + \vect{x}_{2}$. In order to obtain such a result, we choose the matrices $\mat{C}_{1}$ and $\mat{C}_{2}$ such that $\mat{H}_{1,1} \mat{C}_{1} = \mat{H}_{1,2} \mat{C}_{2}$. With appropriate matrices, we get:
\begin{multline}\label{eq:ex_Rx1}
    \vect{Y}_{1} = \mat{H}_{1,1} \mat{C}_{1} \mat{V} (\vect{x}_{1} + \vect{x}_{2}) \\
                + \mat{H}_{1,3} \mat{C}_{3} \mat{V} \vect{x}_{3} + 
                \mat{H}_{1,4} \mat{C}_{4} \mat{V} \vect{x}_{4} + \vect{Z}_{1}
\end{multline}
Based on the received signal at the other receivers, we apply the same selection for the other precoding matrices, we obtain
\begin{multline}\label{eq:ex_Rx2}
    \vect{Y}_{2} = \mat{H}_{2,1} \mat{C}_{1} \mat{V} \vect{x}_{1}  + \mat{H}_{2,2} \mat{C}_{2} \mat{V} (\vect{x}_{2} + \vect{x}_{3}) + \\
                \mat{H}_{2,4} \mat{C}_{4} \mat{V} \vect{x}_{4} + \vect{Z}_{2},
\end{multline}
\begin{multline}\label{eq:ex_Rx3}
    \vect{Y}_{3} = \mat{H}_{3,1} \mat{C}_{1} \mat{V} \vect{x}_{1} + \mat{H}_{3,2} \mat{C}_{2} \mat{V}\vect{x}_{2} \\
                + \mat{H}_{3,3} \mat{C}_{3} \mat{V} (\vect{x}_{3} + \vect{x}_{4}) + \vect{Z}_{3},
\end{multline}
which leads to the following three equations:
\begin{align}\label{eq:cond_AirC1}
    \mat{H}_{1,1} \mat{C}_{1} &= \mat{H}_{1,2} \mat{C}_{2}\\\label{eq:cond_AirC2}
    \mat{H}_{2,2} \mat{C}_{2} &= \mat{H}_{2,3} \mat{C}_{3}\\\label{eq:cond_AirC3}
    \mat{H}_{3,3} \mat{C}_{3} &= \mat{H}_{3,4} \mat{C}_{4}
\end{align}

We get 3 equations for 4 unknowns. We are allowed to choose $\mat{C}_{1}$ randomly and independently from all other matrices, and we can recover the other $\mat{C}_{k}$ by \eqref{eq:cond_AirC1}, \eqref{eq:cond_AirC2} and \eqref{eq:cond_AirC3}: 
\begin{align}\label{eq:def_Ci}
    \mat{C}_{2} &= \mat{H}_{1,2}^{-1} \mat{H}_{1,1} \mat{C}_{1}\\
    \mat{C}_{3} &= \mat{H}_{2,3}^{-1} \mat{H}_{2,2} \mat{H}_{1,2}^{-1} \mat{H}_{1,1}\mat{C}_{1}\\
    \mat{C}_{4} &= \mat{H}_{3,4}^{-1} \mat{H}_{3,3} \mat{H}_{2,3}^{-1} \mat{H}_{2,2} \mat{H}_{1,2}^{-1} \mat{H}_{1,1} \mat{C}_{1}
\end{align}

Then, we construct the IA precoding matrix $\mat{V}$ according to the principle in \cite{cadambe_interference_2008}.
This means that each column of the matrix $\mat V$ is constructed using all the channel matrices that pre-multiply $\mat V$ in the interference terms of the received signals. Each column is associated with a distinct set of exponents applied to these channel matrices. Formally, we choose:

\begin{equation*}
    \mat{V} = \left[ \left(\prod_{\mat{G} \in \mathcal{G}} (\mat G)^{\alpha_{\mat{G}}}\right) \cdot \vect{1} \text{ } : \text{ } \forall \text{ } \boldsymbol{\alpha}_{\mathcal{G}} \in [\![0,n-1]\!]^6 \right],
\end{equation*}
where n is a large number depending of blocklength $\sf T$ that tends to infinity with $\sf T$ and where
\begin{align*}
    \mathcal{G} = \{\mat{H}_{1,3} \mat{C}_{3}, \mat{H}_{1,4} \mat{C}_{4}, \mat{H}_{2,1} \mat{C}_{1}, \\\mat{H}_{2,4} \mat{C}_{4}, \mat{H}_{3,1} \mat{C}_{1}, \mat{H}_{3,4} \mat{C}_{4} \},
\end{align*}
and
$$\boldsymbol{\alpha}_{\mathcal{G}} \triangleq \left( \alpha_{\mat G} \colon \mat G \in \mathcal{G}\right).$$

By this choice of $\mat V$, all interference
signals at Rxs will lie in the column span of the matrix
\begin{equation*}
    \mat{W} = \left[ \left(\prod_{\mat{G} \in \mathcal{G}} (\mat G)^{\alpha_{\mat{G}}}\right) \cdot \vect{1} \text{ } : \text{ } \forall \text{ } \boldsymbol{\alpha}_{\mathcal{G}} \in [\![0,n]\!]^6 \right],
\end{equation*}

The desired signals at the $\ell$-th Rx lie in the subspace spanned by the columns of the matrix 
\begin{equation}
    \mat{A}_{\ell} = [\mat{H}_{\ell,\ell} \mat{C}_{\ell} \mat{V}] \quad \forall \ell \in [\![1,3]\!]
\end{equation}

As proved in \cite{cadambe_interference_2009} and follows from our analysis in next section, with probability 1 (over the random channel matrices) the matrices
\begin{align}
    \mat{\Lambda}_{1} = [\mat{A}_{1}, \mat{W}]
    \\
    \mat{\Lambda}_{2} = [\mat{A}_{2}, \mat{W}]
    \\
    \mat{\Lambda}_{3} = [\mat{A}_{3}, \mat{W}]
\end{align}
have full column-ranks. Therefore, the useful signal can be separated from the interference, allowing to achieve 
\begin{equation*}
    \text{DoF} = \frac{\sf n^{6}}{n^{6} + (n+1)^{6}}
\end{equation*}
for each Rx, which implies that
\begin{equation}
    \text{A-SDoF} = \frac{3}{2}
\end{equation}
in the limit as $n \to \infty$ .

\subsection{General case with $\sf r_{\ell-1, \ell} \leq 1$}\label{sec:proof_general}

Now, we consider the general channel introduced in Section \ref{sec:channel_model}. The network is composed of $\sf K$ clusters and each contains $\sf r$ Txs and one Rx. We recall that the total number of Txs is $\sf M = \sf K \sf r - \sf r_{\sf K}$, where $\sf r_{\sf K}$ represents the number of overlapping clusters in the network. We allow at most one overlap between clusters $k-1$ and $k$, i.e., $\sf r_{k-1,k} \in \{0,1\}$.

We fix a large number $n \in \mathbb{N}$ which we shall let tend to $\infty$ and define
\begin{align}
    \sf \gamma &\triangleq \sf K(\sf M - \sf r)
    \\
    \sf T &\triangleq  n^{\sf \gamma} + \sf (n+1)^{\sf \gamma}
\end{align}
Thus, the blocklength $\sf T$ tends to $\infty$ as $ n$ does. The parameters $\sf \gamma$, $\sf T$ and $n$ are used in the construction of the precoding matrices we shall define later on.

In our scenario, the message $w_{q}$ of Tx $q$ is encoded by $f_{q}$ into a length $\sf T$ vector $\vect{X}_{q}$. We choose the function $f_{q}$ so that we can write
\begin{equation}
    \vect{X}_{q}= \mat{C}_{q} \mat{V} \vect{x}_{q}
\end{equation}
where $\mat{C}_{q}$ is a $\mathbb{C}^{\sf T \times \text{$n$}^{\sf \gamma}}$ precoding matrix, $\mat{V}$ is a $\mathbb{C}^{n^{\sf \gamma} \times n}$ precoding IA matrix and $\vect{x}_{q}$ is a length-$n$ codeword encoding the message $w_{q}$.

According to the channel defined in \eqref{eq:gen}, the received signal at the $\ell$-th Rx can be written as:
\begin{equation}
    \mat{Y}_\ell = \underbrace{\sum_{q \in \TS{\ell}} \mat{H}_{\ell,q} \mat{C}_{q} \mat{V} \vect{x}_q }_\text{useful signals}
    + \underbrace{\sum_{q \in [\![1, \sf M]\!]\backslash \TS{\ell}} \mat{H}_{\ell,q} \mat{C}_q \mat{V} \vect{x}_q}_\text{interfering signals} + \mat{Z}_\ell
\end{equation}

In the considered AirComp scheme, the goal at receiver $\ell$ is to reconstruct the sum of all symbols $\vect{x}_{q}$ for $q \in \TS{\ell}$. To achieve this, precoding matrices $\mat{C}_q$ are designed such that, after going through their respective channels $\mat{H}_{\ell,t}$, all Txs in $\TS{\ell}$ should contribute the same effective signal at Rx $\ell$. This requires the following condition to be satisfied: 
\begin{align}
    \left.
    \begin{array}{c}
    \mat{H}_{\ell,t_{\ell,1}} \mat{C}_{t_{\ell,1}} = \mat{H}_{\ell,t_{\ell,2}} \mat{C}_{t_{\ell,2}} \\
    \mat{H}_{\ell,t_{\ell,1}} \mat{C}_{t_{\ell,1}} = \mat{H}_{\ell,t_{\ell,3}} \mat{C}_{t_{\ell,3}}\\
    \phantom{=}\vdots \\
    \mat{H}_{\ell,t_{\ell,1}} \mat{C}_{t_{\ell,1}} = \mat{H}_{\ell,t_{\ell,\sf r}} \mat{C}_{t_{\ell,\sf r}}\\
    \end{array}
    \right\} \quad \forall \ell \in [\![1,\sf K]\!]
\end{align}
And we recall that $t_{\ell, i}$ is defined as the $i$-th element in the ordered set $\TS{\ell}$.



From the above relations, we can deduce that all matrices $\mat{C}_{t_{\ell, i}}$ with $i \geq 2$ are fully determined by the matrix $\mat{C}_{t_{\ell, 1}}$, i.e.:
\begin{equation}\label{eq:Ci}
    \mat{C}_{q} = \mat{H}_{\ell, q}^{-1} \mat{H}_{\ell, t_{\ell,1}} \mat{C}_{t_{\ell,1}} \quad \forall t_{\ell,1}\leq q \leq t_{\ell, \sf r},\forall \ell \in [\![1,\sf K]\!].
\end{equation}
Notice that, for two adjecting and overlapping Tx groups $\ell$ and $\ell+1$, i.e. $\sf r_{\ell, \ell+1} = 1$, it holds that $t_{\ell,\sf r} = t_{\ell+1, 1}$ since they have a commun Tx. As a result, all the matrices in the two groups are determined recursively from the matrix $\mat{C}_{t_{\ell, 1}}$.
This implies that only the first precoding matrix $\mat{C}_{t_{\ell, 1}}$ from a group that does not overlap with its predecessor needs to be chosen freely; the remaining matrices are then uniquely determined. 
In particular, if each Tx group overlaps with its adjacent groups, selecting $\mat{C}_{1}$ alone suffices to determine all other precoding matrices.
To simplify the notation, we define the set $\mathcal{C}$ containing all such freely chosen precoding matrices:
\begin{equation*}
    \mathcal{C} =\{ \mat{C}_{t_{\ell,1}} |\ell \in [\![1, \sf K]\!] \text{ and } \sf r_{\ell-1,\ell} = 0\}.
\end{equation*}
We choose each matrix in $\mathcal{C}$ to be a diagonal matrix with non-zero entries drawn independently from a continuous distribution. These entries are also independent of all channel matrices and noise.

We now need to design $\mat V$ such that each Rx can successfully decode its intended signals, while the interference is aligned into a small subspace. To ensure this, the interfering signals must not overlap with the subspace occupied by the desired signals.


Inspired by \cite{cadambe_interference_2008} and \cite{bi_wireless_2024}, we design $\mat V$ based on the matrices involved in the interference terms. Specifically, we collect all matrices of the form $\mat{H}_{\ell, q} \mat{C}_q$ that contribute to interference, and use them to construct the columns of $\mat V$. Each column of $\mat V$ is formed as a distinct product of these interference matrices, raised to different integer exponents. Formally, we define:
\begin{equation*}
    \mat{V} = \left[ \left(\prod_{\mat{G} \in \mathcal{G}} (\mat G)^{\alpha_{\mat{G}}}\right) \cdot \vect{1} \text{ } : \text{ } \forall \text{ } \boldsymbol{\alpha}_{\mathcal{G}} \in [\![0,n-1]\!]^{\sf \gamma} \right],
\end{equation*}
where $\mathcal{G}$ contains all the matrices of the interfering signals:
\begin{equation*}
    \mathcal{G} = \{\mat{H}_{\ell,k} \mat{C}_{k} \mid \ell \in [ \![1,\sf K]\!] \text{ and $k$} \in[ \![1,\sf M]\!]\setminus \TS{\ell} \}
\end{equation*}
and
$$\boldsymbol{\alpha}_{\mathcal{G}} \triangleq \left( \alpha_{\mat G} \colon \mat G \in \mathcal{G}\right).$$

Since the space spanned by the columns of $\mat{V}$ contains all power products of powers $1$ to $n-1$ of the matrices $\mat{G} \in \mathcal{G}$, we have
\begin{equation}
    span(\mat{G} \cdot \mat{V}) \subset span(\mat{W}) \quad \forall \mat{G} \in \mathcal{G}
\end{equation}
where $span(\mat{B})$ denotes the space spanned by the column of the matrix $\mat{B}$, and we defined the $\sf T \times (\text{$n$} + 1)^{\sf \gamma}$ matrix
\begin{equation*}
    \mat{W} = \left[ \left(\prod_{\mat{G} \in \mathcal{G}} (\mat G)^{\alpha_{\mat{G}}}\right) \cdot \vect{1} \text{ } : \text{ } \forall \text{ } \boldsymbol{\alpha}_{\mathcal{G}} \in [\![0,n]\!]^{\sf \gamma} \right].
\end{equation*}

The signal and interference space at Rx $\ell$ is represented by the matrix:
\begin{equation}
    \mat{\Lambda}_{\ell} = [\mat{H}_{\ell,t_{\ell,1}}\mat{C}_{t_{\ell,1}}\mat{V}, \quad \mat{W}]
\end{equation}
The first part of $\mat{\Lambda}_{\ell}$ represent the useful signal subspace and consists of a matrix of dimension $\sf T \times \text{$n^{\sf \gamma}$}$. Since the interference space is represented by the matrix $\mat{W}$ which is of dimension $\sf T \times \text{$(n+1)^{\sf \gamma}$}$, received matrix $\mat{\Lambda}_{\ell}$ is square $\sf T \times \sf T$.
To prove the full rankness of the matrix $\mat{\Lambda}_{\ell}$, we introduce the following lemma.

\begin{lemma}\label{lemma:full-rank}
    Let $\vect{s}_1$, $\vect{s}_2$, ..., $\vect{s}_L$ be independent random vectors with i.i.d. entries drawn according to continuous distributions. Let $L$ different exponent vectors
    \begin{equation*}
        \vect{\alpha}_{j} = (\alpha_{j,1}, \cdots, \alpha_{j,L}) \in \mathbb{N}^{L} \quad j \in [\![1,L]\!],
    \end{equation*}
    the $L \times L$ matrix $\vect{M}$ with row-$i$ and column-$j$ entry 
    \begin{equation*}
        M(i,j) = \prod_{k=1}^{L} ({s}_{k}(i))^{\alpha_{j,k}}, \qquad i \in [\![1,L]\!], j \in [\![1,L]\!]
    \end{equation*}
    is full rank almost surely.
\end{lemma}

\lo{\begin{IEEEproof} 
The main idea of the proof is as follows:

We view $\det(\mat{M})$ as a multivariate polynomial in the first-row variables $s_k(1)$, with coefficients depending on the remaining entries. If $\det(\mat{M}) = 0$ with nonzero probability, then either it is the zero polynomial or its variables must lie in the zero set of a nonzero polynomial, which has measure zero. Iterating this reasoning on all minors shows that the zero determinant would imply all entries are zero with positive probability, contradicting the assumption of continuous distributions.
Hence, $\det(\mat{M}) \neq 0$ almost surely, and $\mat{M}$ is of full rank with probability one.
For further details, see Appendix~\ref{sec:app_lemma_full-rank} \end{IEEEproof}}
\gc{
\begin{IEEEproof}
See \cite{cadambe_interference_2009}.
\end{IEEEproof}
}

In order to apply Lemma~\ref{lemma:full-rank}, we rewrite column in the matrix $\Lambda_\ell$ into the following form: 
\begin{equation}\label{eq:col_V}
    \left(\prod_{\mat{H} \in \mathcal{H}_{1}} (\mat H)^{\boldsymbol{\alpha}_{\mat{H}}}\right) \cdot 
    \left(\prod_{\mat{H'} \in \mathcal{H}_{2}} (\mat H')^{\alpha_{\mat{H'}}}\right) \cdot
    \left(\prod_{\mat{C} \in \mathcal{C}} (\mat C)^{\alpha_{\mat{C}}}\right) \cdot
    \vect{1}
\end{equation}
where
\begin{align*}
    \mathcal{H}_{1} &= \{ \mat{H}_{\ell,q} \mid \ell \in [ \![1,\sf K]\!] \text{ and } \text{$q$} \in[ \![1,\sf M]\!] 
    \setminus \TS{\ell} \}
    \\
    \mathcal{H}_{2} &= \{ \mat{H}_{\ell, q} \mid \forall \ell \in [\![1,\sf K]\!], \forall \text{$q$} \in \TS{\ell} \}
\end{align*}
$\mathcal{H}_{1}$ is the set of channel matrices from the interfering signals and $\mathcal{H}_{2}$ is the set of those from the useful signals. We notice that all channel matrices used to construct $\mathcal{H}_{1}$ and $\mathcal{H}_{2}$ are distinct.
As a result, any two matrices selected from the union $\mathcal{H}_{1} \cup \mathcal{H}_{2} \cup \mathcal{C}$ are independent, which corresponds to vectors $s$ in Lemma~\ref{lemma:full-rank}. To prove that the matrix $\Lambda_\ell$ is full rank, it therefore suffices to show that each of its columns has a unique exponent vector.

We select two different columns $\vect{c}_1$ and $\vect{c}_2$ from the matrix $\Lambda_\ell$ with $\left(\balpha_{\mat H}^{(1)}, \balpha_{\mat H'}^{(1)}, \balpha_{\mat C}^{(1)}\right)$ and $\left(\balpha_{\mat H}^{(2)}, \balpha_{\mat H'}^{(2)}, \balpha_{\mat C}^{(2)}\right)$. There are three cases:
\begin{itemize}
    \item \textbf{Both $\vect{c}_1$ and $\vect{c}_2$ are selected from the matrix $\mat W$.} In this case, the corresponding exponent vectors are $\balpha_{\mat G}^{(1)}$ and $\balpha_{\mat G}^{(2)}$.
    Note that each matrix in $\mathcal{G}$ is a product of two matrices: $\mat H$ and $\mat C$. Specifically, $\mat H$ belongs to $\mathcal{H}_1$, while $\mat C$ is itself a product of matrices only from $\mathcal{H}_2$ and $\mathcal{C}$, as shown in \eqref{eq:Ci}.
    Since $\mat H$ and $\mat C$ are built from disjoint sets $\mathcal{H}_1$ and $\mathcal{H}_2$, we have $\balpha_{\mat G} = \balpha_{\mat H}$.
    Therefore, the exponent vectors $\balpha_{\mat H}^{(1)}$ and $\balpha_{\mat H}^{(2)}$ are different.

    \item \textbf{Both $\vect{c}_1$ and $\vect{c}_2$ are selected from the useful signal subspace.}
    The same reasoning applies as in the previous case. The difference in exponent vectors implies that $\balpha_{\mat H}^{(1)} \ne \balpha_{\mat H}^{(2)}$.

    \item \textbf{$\vect{c}_1$ is selected from $\mat W$, and $\vect{c}_2$ is selected from $\mat V$.}
    In this case, we focus on the exponent vector $\balpha_{\mat H'}$. The only factors contributing to $\balpha_{\mat H'}$ come from $\mat{C}_q$ in $\mathcal{G}$, and $\mat{C}_q = \mat{H}_q^{-1}\mat{H}_{t_{\ell,1}}\mat{C}_{t_{\ell,1}}$. This product consists of two matrices from $\mathcal{H}_2$ with opposite exponents, and a matrix from $\mathcal{C}$. Therefore, as $\vect{c}_1$ is selected from $\mat W$, the net contribution to $\balpha_{\mat H'}^{(1)}$ from the $\mathcal{H}_2$ matrices is zero. 
    On the other hand, as $\vect{c}_2$ is selected from the signal subspace, it is a columns of the form $\mat{H}_{\ell,t_{\ell,1}}\mat{C}_{t_{\ell,1}}\mat{V}$, where $\mat{H}_{\ell,t_{\ell,1}} \in \mathcal{H}_2$. This results in a total contribution of one to the sum of the components in $\balpha_{\mat H'}^{(2)}$.
\end{itemize}
We thus conclude that each column has different exponent vector $\left(\balpha_{\mat H}, \balpha_{\mat H'}, \balpha_{\mat C}\right)$, which imply that matrices $\{\mat{\Lambda}_{\ell}\}$ are full column rank almost surely.

This proves that each Rx $\ell$ can separate the various desired signals from each other as well as from the non-desired interfering signals. And since each receiver occupies $n^{\sf \gamma}$ dimension out of $\sf T$, we obtain that a computation rate $\sf R_\ell$ can be achieved such that
\begin{equation}
    \lim_{\sf P \to \infty} \frac{\sf R_\ell}{\sf R} = \frac{n^{\sf \gamma}}{\sf T}.
\end{equation}
with arbitrarily small probability of error as $n \to \infty$.
Since 
\begin{equation}
    \lim_{n \to \infty} \frac{n^{\sf \gamma}}{\sf T} = \frac{1}{2},
\end{equation}
we conclude that 
\begin{equation}
    \text{A-SDoF} = \frac{\sf K}{2}
\end{equation}
is achievable over the system, which conclude the proof of Theorem~\ref{theorem1}. 

\gc{
The second part of the theorem addresses the case where two adjacent groups share more than one Tx. The analysis follows a similar framework, but requires the use of two distinct precoding matrices, $\mat{V}_1$ and $\mat{V}_2$. Specifically, codewords intended for receivers with odd indices are precoded with $\mat{V}_1$, while those intended for receivers with even indices are precoded with $\mat{V}_2$. The complete proof can be found in our ArXiv version \cite{}.
}

\lo{
\subsection{General case with $\sf r_{\ell-1, \ell} \geq 2$}
\yue{In this subsection, we allow the two adjacent groups to share more than one Tx. Thus, we allow $\sf r_{\ell-1,\ell}+r_{\ell,\ell+1} \leq \sf r$.} All other parameters are as in the previous subsection, except that we set 
$$\sf T = n^{\gamma} + 2(n+1)^{\gamma}.$$ 
\begin{remark}
$\sf T$ is larger because multiple Txs may now serve two Rxs simultaneously, making the scheme from the previous subsection inapplicable. For example, suppose clusters $1$ and $2$ share two Tx $r-1$ and Tx $r$. Both Txs would then send messages to Rxs $1$ and $2$, leading to the conditions
\begin{align}
    \mat{H}_{1,\sf r-1} \mat{C}_{\sf r-1} = \mat{H}_{1,\sf r} \mat{C}_{\sf r}\\
    \mat{H}_{2,\sf r-1} \mat{C}_{\sf r-1} = \mat{H}_{2,\sf r} \mat{C}_{\sf r}
\end{align}
which have no solution other than the zero matrix. Indeed, more than $1$ overlap increases the number of equations. Therefore, we need to adapt our strategy. 
\end{remark}

\yue{We apply two different precoding matrices $\mat{V}_1$ and $\mat{V}_2$. Codewords intended for receivers with \textbf{odd indices are assigned the precoding matrix} $\mat{V}_1$, while those intended for receivers with \textbf{even indices are assigned} $\mat{V}_2$. Therefore, the output signal of Tx $q \in \TS{\ell}$ with $\ell$ even is
\begin{equation}
    \mat{X}_{q} = 
    \left\{
    \begin{aligned}
    & \left(\mat{C}_{\ell-1}^{(q)} \mat{V}_{1} + \mat{C}_{\ell}^{(q)} \mat{V}_{2} \right)\vect{x}_{q}, \quad \text{if $q\in \TS{\ell-1}$} \nonumber \\
    & \left(\mat{C}_{\ell}^{(q)} \mat{V}_{2} + \mat{C}_{\ell+1}^{(q)} \mat{V}_{1} \right) \vect{x}_{q}, \quad \text{if $q\in \TS{\ell+1}$} \nonumber \\
    & \mat{C}_{\ell}^{(q)} \mat{V}_{2} \vect{x}_{q} \quad \text{otherwise}
    \end{aligned}
    \right.
\end{equation}
where $\mat{C}_{\ell}^{(q)}$ is a $\mathbb{C}^{\sf T \times n^{\gamma}}$ precoding matrix, $\mat{V}_{i}$ is a $\mathbb{C}^{n^{\gamma} \times n}$ precoding IA matrix, and the construction of these precoding matrices will be discussed shortly. $\vect{x}_{q}$ is a length-n codeword encoding the data $\psi(d_{q})$. Similarly, the output signal of Tx $q \in \TS{\ell}$ with $\ell$ odd is
\begin{equation}
    \mat{X}_{q} = 
    \left\{
    \begin{aligned}
    & \left(\mat{C}_{\ell-1}^{(q)} \mat{V}_{2} + \mat{C}_{\ell}^{(q)} \mat{V}_{1} \right)\vect{x}_{q}, \quad \text{if $q\in \TS{\ell-1}$} \nonumber \\
    & \left(\mat{C}_{\ell}^{(q)} \mat{V}_{1} + \mat{C}_{\ell+1}^{(q)} \mat{V}_{2} \right) \vect{x}_{q}, \quad \text{if $q\in \TS{\ell+1}$} \nonumber \\
    & \mat{C}_{\ell}^{(q)} \mat{V}_{1} \vect{x}_{q} \quad \text{otherwise}
    \end{aligned}
    \right.
\end{equation}
}

For the scheme has changed, the goal is still the same, reconstruct the sum of all symbols $\vect{x}_{q}$ for $q \in \TS{\ell}$. Thus, we get the following condition based on our previous work:
\begin{align}
    \left.
    \begin{array}{c}
    \mat{H}_{\ell, t_{\ell,1}} \mat{C}_{\ell}^{(t_{\ell,1})} = \mat{H}_{\ell, t_{\ell,2}} \mat{C}_{\ell}^{(t_{\ell,2})} \\
    \vdots \\
    \mat{H}_{\ell, t_{\ell,1}} \mat{C}_{\ell}^{(t_{\ell,1})} = \mat{H}_{\ell, t_{\ell,\sf r}} \mat{C}_{\ell}^{(t_{\ell,\sf r})} \\
    \end{array}
    \right\} \quad \forall \ell \in [\![1,\sf K]\!]
\end{align}
As before, we can deduce that all matrices $\mat{C}^{(t_{\ell,i})}_{\ell}$ with $i \geq 2$ are fully determined by the matrix $\mat{C}^{(t_{\ell,1})}_{\ell}$ as in \eqref{eq:Ci}.
It is important to remember that, from our setting, the overlapping Txs have $2$ different messages to send, depending on the Rx. Since the pre-encoding introduce now two matrices $\mat{C}_{\ell}^{(q)}$ and $\mat{C}_{\ell'}^{(q)}$ for each receiver, the matrices cannot be determined recursively from a group to another. This implies that the first precoding matrices from each group needs to be chosen freely. 

The number of IA precoding matrices also differ in this scenario. We need to design two special matrix that aligned the interference in a subspace for the two messages sent by the same transmitter. We define:
\begin{IEEEeqnarray}{rCl}
    \mat{V}_{i} &=& \left[ \Bigg( \prod_{\mat{G} \in \mathcal{G}_{i}}(\mat{G})^{\alpha_{\mat{G}}} \Bigg) \cdot \vect{\Xi}_{i} : \forall \vect{\alpha}_{\mathcal{G}_{i}} \in [\![0, n-1]\!]^{\gamma} \right], \nonumber\\ && \hspace{3.5cm} \text{ for } i \in \{1,2\},
\end{IEEEeqnarray}
where $\mathcal{G}_{i}$ contains all the matrices of the interfering signals:
\begin{equation*}
    \mathcal{G}_{1} = \{ \mat{H}_{\ell, k} \mat{C}_{\ell}^{(k)} \mid \ell \in [\![1, \sf K]\!], \text{odd, and } k \in [\![1, \sf M]\!] \setminus \TS{\ell} \},
\end{equation*}
\begin{equation*}
    \mathcal{G}_{2} = \{ \mat{H}_{\ell, k} \mat{C}_{\ell}^{(k)} \mid \ell \in [\![1, \sf K]\!], \text{ even, and } k \in [\![1, \sf M]\!] \setminus \TS{\ell} \},
\end{equation*}
\begin{equation*}
    \vect{\alpha}_{\mathcal{G}_{i}} \triangleq (\alpha_{\mat{G}} : \mat{G} \in \mathcal{G}_{i})
\end{equation*}
and $\vect{\Xi}_{i}$ are i.i.d. random vectors independent of all channel matrices, noises, and messages as explained in \cite{bi_wireless_2024}.

The choice of $\mat{V}_i$ ensure that the subspace of the interfering signals lies within the column span of the matrix $[ \mat{W}_1, \mat{W}_2]$ with
\begin{IEEEeqnarray}{rCl}
    \mat{W}_{i} &=& \left[ \Bigg( \prod_{\mat{G} \in \mathcal{G}_{i}}(\mat{G})^{\alpha_{\mat{G}}} \Bigg) \cdot \vect{\Xi}_{i} : \forall \vect{\alpha}_{\mathcal{G}_{i}} \in [\![0, n-1]\!]^{\gamma} \right], \nonumber\\ && \hspace{3.5cm} \text{ for } i \in \{1,2\}.
\end{IEEEeqnarray}
We consider the case with $\ell$ odd (the same analysis can be done with $\ell$ odd), the signals and interference at the $\ell$-th Rx lie in the subspace spanned by the columns of the matrix
\begin{equation}
    \mat{\Lambda}_{\ell} = [\mat{H}_{\ell,t_{\ell,1}} \mat{C}_{\ell}^{(t_{\ell,1})} \mat{V}_{1}, \mat{W}_{1}, \mat{W}_{2}]
\end{equation}
To prove that the matrix $\Lambda_\ell$ is of full rank, we only need to show that each of its columns has a unique exponent vector. We select two different columns $\vect{c}_1$ and $\vect{c}_2$ from the matrix. If the two columns are from the signal space and $\mat{W}_1$, or are only from $\mat{W}_2$, the same argument of the three cases towards the end of Section \ref{sec:proof_general} can be applied. Otherwise, it is obvious that they have different exponents, as one has factor $\Xi_1$, while the other has factor $\Xi_2$.

To prove that the matrix $\Lambda_\ell$ is of full rank, it suffices to show that each column has a distinct exponent vector. Consider two different columns, $\vect{c}_1$ and $\vect{c}_2$. If both columns come from the signal space and $\mat{W}_1$, or both from $\mat{W}_2$, then we can apply the same reasoning as in the three cases discussed at the end of Section~\ref{sec:proof_general}. In all other cases, the columns clearly have different exponents, since one involves the factor $\Xi_1$ while the other involves $\Xi_2$.

We conclude that the network achieves 
\begin{equation*}
    \text{DoF} = \frac{1}{3}
\end{equation*}
and so 
\begin{equation}
    \text{A-SDoF} = \frac{\sf K}{3}
\end{equation}
}


%% file: contents/Lemma_full_rank.tex
\section{Proof Lemma~\ref{lemma:full-rank}}\label{sec:app_lemma_full-rank}

Let $\vect{s}_1$, $\vect{s}_2$, ..., $\vect{s}_L$ be independent random vectors with i.i.d. entries drawn according to continuous distributions. Lets intorduce $L$ different exponent vectors
\begin{equation*}
    \vect{\alpha}_{j} = (\alpha_{j,1}, \cdots, \alpha_{j,L}) \in \mathbb{N}^{L} \quad j \in [\![1,L]\!],
\end{equation*}
define the $L \times L$ matrix $\mat{M}$ with row-$i$ and column-$j$ entry
\begin{equation*}
    M(i,j) = \prod_{k=1}^{L}(s_{k}(i))^{\alpha_{j,k}}, \quad i,j \in [\![1,L]\!].
\end{equation*}

We need to show that the determinant of $\mat{M}$ is non zero with probability $1$. Let introduce $\mat{C}_{i,j}$ the cofactor of $M(i,j)$. 
\begin{equation*}
    \text{det}(\mat{M}) = \mat{C}_{1,1}M(1,1) + \cdots + \mat{C}_{1,L}M(1,L)
\end{equation*}
Since $M(1,j)$ is a product of $s_{k}(1),$ $k \in [\![1,L]\!]$ put at a certain power, it implies that we can see $\text{det}(\mat{M})$ as a polynomial in $s_{k}(1),$ which coefficients are the $C_{1,j}$. Therefore, $\text{det}(\mat{M}) = 0$ iff the $s_{k}(1),$ $k \in [\![1,L]\!]$ are the roots of this polynomial, or it is the zero polynomial.

By contradiction, assume $\text{det}(\mat{M}) = 0$ with non zero probability. We also assume it's not because of the zero polynomial, which means at least one of the $C_{1,j}$ coefficients are non zero. Then, the set of roots is a finite set. We do have to point out that the $C_{1,j}$ coefficients are function of $s_{k}(\ell),$ $k, \ell \in [\![1,L]\!]$. Therefore, the variables $s_{k}(\ell)$ are drawn independently on a continuous distributions conditioned by all the $C_{1,j}$. But the probability their are taken within a finite set is still $0$. Therefore, the polynomial we were looking at must be the zero polynomial.

Thus, $C_{1,j} = 0$ for all $j \in [\![1,L]\!]$. It implies that all the minors $\text{det}(\mat{\widetilde{M}}(1,j)) = 0$ with non zero probability, where $\mat{\widetilde{M}}(1,j)$ stands for the same matrix $\mat{M}$ but without the first row and $j$-th column. Therefore, the same reasoning can be applied once again but on every sub matrices $\mat{\widetilde{M}}(1,j)$ which leads to the conclusion that their cofactors are all zero. We can iterate this process until we get $1 \times 1$ matrices which coefficients are again $s_{k}(\ell),$ $k, \ell \in [\![1,L]\!]$. Since they are drawn independently on a continuous distributions, the probability that they are zero is $0$. This is in contradiction with our first assumption.

Therefore, $\text{det}(\mat{M}) = 0$ with zero probability. So $\mat{M}$ is full rank almost surely.

%% file: conference_itw2025.bbl
\begin{thebibliography}{10}
\providecommand{\url}[1]{#1}
\csname url@samestyle\endcsname
\providecommand{\newblock}{\relax}
\providecommand{\bibinfo}[2]{#2}
\providecommand{\BIBentrySTDinterwordspacing}{\spaceskip=0pt\relax}
\providecommand{\BIBentryALTinterwordstretchfactor}{4}
\providecommand{\BIBentryALTinterwordspacing}{\spaceskip=\fontdimen2\font plus
\BIBentryALTinterwordstretchfactor\fontdimen3\font minus \fontdimen4\font\relax}
\providecommand{\BIBforeignlanguage}[2]{{%
\expandafter\ifx\csname l@#1\endcsname\relax
\typeout{** WARNING: IEEEtran.bst: No hyphenation pattern has been}%
\typeout{** loaded for the language `#1'. Using the pattern for}%
\typeout{** the default language instead.}%
\else
\language=\csname l@#1\endcsname
\fi
#2}}
\providecommand{\BIBdecl}{\relax}
\BIBdecl

\bibitem{zhu_one_bit_aggregation_2021}
G.~Zhu, Y.~Du, D.~Gündüz, and K.~Huang, ``One-bit over-the-air aggregation for communication-efficient federated edge learning: Design and convergence analysis,'' \emph{IEEE Transactions on Wireless Communications}, vol.~20, no.~3, pp. 2120--2135, Mar. 2021.

\bibitem{zhu_mimo_aircomp_2019}
G.~Zhu and K.~Huang, ``Mimo over-the-air computation for high-mobility multimodal sensing,'' \emph{IEEE Internet of Things Journal}, vol.~6, no.~4, pp. 6089--6103, Aug. 2019.

\bibitem{zhu_aircomp_aggregation_2021}
G.~Zhu, J.~Xu, K.~Huang, and S.~Cui, ``Over-the-air computing for wireless data aggregation in massive iot,'' \emph{IEEE Wireless Communications}, vol.~28, no.~4, pp. 57--65, Aug. 2021.

\bibitem{wen_aircomp_2019}
D.~Wen, G.~Zhu, and K.~Huang, ``Reduced-dimension design of mimo over-the-air computing for data aggregation in clustered iot networks,'' \emph{IEEE Transactions on Wireless Communications}, vol.~18, no.~11, pp. 5255--5268, Nov. 2019.

\bibitem{li_aircomp_2019}
X.~Li, G.~Zhu, Y.~Gong, and K.~Huang, ``Wirelessly powered data aggregation for iot via over-the-air function computation: Beamforming and power control,'' \emph{IEEE Transactions on Wireless Communications}, vol.~18, no.~7, pp. 3437--3452, Jul. 2019.

\bibitem{jafar_interference_2011}
S.~A. Jafar, \emph{Interference Alignment: A New Look at Signal Dimensions in a Communication Network}.\hskip 1em plus 0.5em minus 0.4em\relax Foundations and Trends® in Communications and Information Theory, 2011.

\bibitem{cadambe_interference_2008}
V.~R. Cadambe and S.~A. Jafar, ``Interference alignment and degrees of freedom of the {K}-user interference channel,'' \emph{IEEE Transactions on Information Theory}, vol.~54, no.~8, pp. 3425--3441, Aug. 2008.

\bibitem{cadambe_interference_2009}
------, ``Interference alignment and the degrees of freedom of wireless x networks,'' \emph{IEEE Transactions on Information Theory}, vol.~55, no.~9, pp. 3893--3908, Sep. 2009.

\bibitem{annapureddy_DoF_2012}
V.~S. Annapureddy, A.~El~Gamal, and V.~V. Veeravalli, ``Degrees of freedom of interference channels with comp transmission and reception,'' \emph{IEEE Transactions on Information Theory}, vol.~58, no.~9, pp. 5740--5760, Sep. 2012.

\bibitem{bi_DoF_2022}
Y.~Bi, P.~Ciblat, M.~Wigger, and Y.~Wu, ``Dof of a cooperative x-channel with an application to distributed computing,'' in \emph{2022 IEEE International Symposium on Information Theory (ISIT)}, Jun. 2022, pp. 566--571.

\bibitem{tang_interference_2013}
J.~Tang and S.~Lambotharan, ``Interference alignment techniques for mimo multi-cell interfering broadcast channels,'' \emph{IEEE Transactions on Communications}, vol.~61, no.~1, pp. 164--175, Jan. 2013.

\bibitem{mohammadghasemi_feedback_2019}
H.~Mohammadghasemi, M.~F. Sabahi, and A.~R. Forouzan, ``Limited feedback distributed interference alignment in cellular networks with large scale antennas,'' \emph{AEU - International Journal of Electronics and Communications}, vol. 110, Oct. 2019.

\bibitem{lan_combined_2020}
Q.~Lan, H.~S. Kang, and K.~Huang, ``Simultaneous signal-and-interference alignment for two-cell over-the-air computation,'' \emph{IEEE Wireless Communications Letters}, vol.~9, no.~9, pp. 1342--1345, Sep. 2020.

\bibitem{li_interference_2024}
S.~Li, M.~Sun, X.~Cui, and J.~Liu, ``Channel reconfiguration distributed interference alignment for multi-unit over-the-air computation,'' \emph{AEU - International Journal of Electronics and Communications}, vol. 175, Feb. 2024.

\bibitem{bi_wireless_2024}
Y.~Bi, M.~Wigger, and Y.~Wu, ``Normalized delivery time of wireless mapreduce,'' \emph{IEEE Transactions on Information Theory}, vol.~70, no.~10, pp. 7005--7022, Oct. 2024.

\bibitem{Sahin_survey_AirComp_2023}
A.~Şahin and R.~Yang, ``A survey on over-the-air computation,'' \emph{IEEE Communications Surveys \& Tutorials}, vol.~25, no.~3, pp. 1877--1908, 2023.

\bibitem{Nazer_compute_2011}
B.~Nazer and M.~Gastpar, ``Compute-and-forward: Harnessing interference through structured codes,'' \emph{IEEE Transactions on Information Theory}, vol.~57, no.~10, pp. 6463--6486, 2011.

\bibitem{Jeon_computation_2014}
S.-W. Jeon, C.-Y. Wang, and M.~Gastpar, ``Computation over gaussian networks with orthogonal components,'' \emph{IEEE Transactions on Information Theory}, vol.~60, no.~12, pp. 7841--7861, 2014.

\bibitem{Nomo_functions_lattices}
M.~Goldenbaum, H.~Boche, and S.~Stańczak, ``Nomographic functions: Efficient computation in clustered gaussian sensor networks,'' \emph{IEEE Transactions on Wireless Communications}, vol.~14, no.~4, pp. 2093--2105, 2015.

\end{thebibliography}
